

\magnification=1200
\baselineskip=20pt
\ \
\def\cl{\centerline}

\def\Mo{M_{\odot}}
\hyphenation{Schwarz-schild}

\vskip 5.0 cm
\cl{\bf FINITE SOURCE SIZES AND THE INFORMATION CONTENT OF }
\cl{\bf MACHO-TYPE LENS SEARCH LIGHT CURVES }
\vskip 3 cm
\baselineskip=15pt
\cl { Robert J. Nemiroff$^{1,2}$ and W. A. D. T. Wickramasinghe$^{3}$ }
\bigskip
\bigskip\bigskip
\cl{$^1$ George Mason University, CSI, Fairfax, VA 22030 }
\cl{$^2$ NASA / Goddard Space Flight Center, Code 668.1, Greenbelt, MD 20771 }
\cl{$^3$ University of Pennsylvania, Philadelphia, PA 19104 }
\bigskip\bigskip
\bigskip\bigskip

\cl{ In press: }

\cl{ Astrophysical Journal (Letters) }

\vfill\eject

\baselineskip=20pt
\parindent=20pt
\parskip=5pt

\cl{\bf ABSTRACT }

If the dark halo matter is primarily composed of MACHOs toward the lower
end of the possible detection range ($ < 10^{-3}$ $\Mo$) a fraction of the
lens detection events should involve the lens crossing directly in front of
the disk of the background star. Previously, Nemiroff (1987) has shown that
each crossing would create an inflection point in the light curve of the
MACHO event.  Such inflection points would allow a measure of the time it
took for the lens to cross the stellar disk. Given an independent estimate
of the stellar radius by other methods, one could then obtain a more
accurate estimate of the velocity of the lens. This velocity could then, in
turn, be used to obtain a more accurate estimate of the mass range for the
MACHO or disk star doing the lensing.

\noindent
{\it Subject Headings:} stars: low-mass, Galaxy: halo, dark matter,
gravitational lensing

\vfill\eject

\cl{\bf 1. Introduction }

Recently Alcock et al. (1993, the MACHO collaboration, where MACHO stands
for Massive Compact Halo Object), Aubourg et al. (1993, the Experience de
Recherche d'Objets Sombres or EROS project) and Udalski et al. (1993,
called Optical Gravitational Lensing Experiment or OGLE) have all reported
seeing light curves of stars indicative of a fainter star passing in front
of and gravitationally magnifying the light from a background star: a
gravitational lens event.  The probability of seeing such an event was
predicted originally by Paczynski (1986) and by Griest (1991), while an
estimate fully including relative lens and source motion was given by
Nemiroff (1991).  Assuming that the nature of these events is correctly
identified, these events are measuring the mass and density of stars in our
disk and dark-matter halo objects in the Galactic halo.

Paramount to the success of these efforts is the ability to turn light
curves into useful information about the mass and density of the lenses. In
this paper we use the fact that a reasonable fraction of strong
magnification lens events involving low mass MACHOs ($< 10^{-3}$ $\Mo$)
would involve the lens crossing the finite stellar disk of the background
star (Witt and Mao 1994).  In general, the smaller the mass of the MACHO,
the more of them are needed to explain the rotation curve of our Galaxy,
the more likely one will cross directly in front of a background stellar
disk.  Also the smaller the mass of the lens, the smaller the angular size
of its Einstein ring relative to the angular size of a background stellar
disk, the more likely that large magnitude lensing events will involve a
disk crossing.  For these reasons low mass MACHOs are considered
to be the most probable lenses for disk crossing events.

Lensing effects on a finite-sized source were discussed previously by
Nemiroff (1987), by Schneider and Wagoner (1987) in the context of
analyzing gravitational lensing effects of distant supernovae, and more
generally in the book on gravitational lenses by Schneider, Ehlers and
Falco (1992).  Gould (1992) discussed the logistics of detecting objects as
low as $10^{-9}$ $\Mo$ and gave a description of the shape of a light curve
for a lens crossing a stellar disk.

Clearly a gravitational lensing light curve will become more complex when
finite source sizes are included in the lensing scenario, as shown in
Nemiroff (1987).  Although this may be thought of as unfortunate, since it
makes understanding the light curves more complicated, the added
information available in the light curve will be shown to be useful. In the
next section the information content in point and finite source size MACHO
lens events is discussed with a goal of using this extra information to
better deconvolve the mass and relative velocity of the lens.  The last
section gives a summary and some discussion.

\bigskip

\cl{\bf 2. The Information Content of Point Source MACHO Events }

If a lens passes in front of a background source which is considered to be
a point, one parameter completely describes the shape of the light curve:
the angular impact parameter between the lens and the source ($B$). A
second parameter acts like a multiplier in the duration of the light
curve: the relative angular speed of the lens across the field containing
the source which we will designate $V$. Yet a third parameter must be used
to locate the time of light curve maximum, but we will assume that this
zero-point temporal orientation is unambiguous here.

Both $B$ and $V$ are only determined from the light curve in terms of the
projected angular Einstein ring size of the lens. More precisely, the
measurable angular impact parameter
 $$ B = { \beta \over E } ,
 \eqno(1)$$
where $\beta$ is the angular impact parameter between the lens and the
source (the closest angular approach of the lens from the source) and $E$
is the angular size of the Einstein ring of the lens, and is given by
(Liebes 1964, Refsdal 1964)
 $$ E = \sqrt{ 2 R_S (D_{OS} - D_{OL}) \over
               D_{OL} D_{OS} } ,
 \eqno(2)$$
where $D_{OL}$ is the distance between the observer and the lens, $D_{OS}$
is the distance between the observer and the source, and $R_S$ is the
Schwarzschild radius of the lens (related to mass by $R_S = 3 {\rm km} \,
(M/M_{\odot}$).  $B$ is related to the maximum magnification of lensing $A$
by
  $$ B = \left[ 2 \sqrt{ A^2 \over A^2 - 1 } - 2 \right]^{1/2} ,
 \eqno(3)$$
such that for large $A$ (small $B$), they are simply related by $B \approx
1/A$.  Note that Eq. 3 holds even if $B$ is interpreted as any relative
projected distance of the lens from the source, not only the minimum
distance (impact parameter) as used here.

The measurable relative angular velocity parameter between the lens and the
source which can be determined from the light curve is
 $$ V = { v / E } ,
 \eqno(4)$$
where $v$ is the actual angular velocity of lens relative to the source in
the lens plane.

The most information a point source light curve can hope to provide is,
through a well determined shape, accurate determinations of $B$, $V$ and
the time of maximum light. Once they are determined, one must assume a lens
distance $D_{OL}$, a source distance $D_{OS}$, and a projected relative
angular lens velocity $v$ in order to solve for the mass of the lens.  (If
the source distance is much greater than the lens distance then the source
distance is not important.) Relative angular lens velocities are
particularly unconstrained as there is usually no indication what type of
orbit the lens or source is on, so that its velocity may be uncertain by an
order of magnitude.  This uncertainly translates directly to uncertainty in
the mass of the lens.

More information is discernable from the light curve of a source which has
a finite angular size.  Specifically, a parameter involving the size of the
source is recoverable. Generally, finite size sources are only important if
the angular size of the source is comparable with $B$.  If $B$ is much
larger, light curves of point and finite sources will be practically
indistinguishable.

If the MACHO crosses the disk of the source star, Nemiroff (1987) has found
that lens crossing is closely matched in time with an inflection point of
the light curve. This is reproduced here as Figure 1.  This effect is also
discernable from Figure 11.1 of Schneider, Ehlers, and Falco (1992).
Therefore source sizes are relatively easy to discern from the lens
deconvolution analysis of the light curve. Information is not lost from the
previous point-source case - both $B$ and $V$ values that would have been
measured in the absence of finite source size can be determined by fits to
the light curve far from the peak, where the finite source size is not
important. But now, however, the crossing time of the lens in front of the
source can be determined as
 $$ T \sim { (R_*/D_{OS}) \over (v/D_{OL}) } ,
 \eqno(5)$$
where $T$ is the measured time between lens crossings of the source disk,
as determined by noting the time of inflection points, and $R_*$ is the
physical radius of the star.  $T$ is directly measured from the light
curve, and $R_*$ can be estimated independently (by noting the stellar
type) for the source star.  This allows one to independently estimate $v$,
the projected angular velocity between the lens and the source. Possible
more importantly, one can then use Equation (4) to compute $E$, the angular
Einstein ring size, and then compute $R_S$ and hence a more accurate mass
of the lens through Equation (2).

For a source star with $R_* = 30 R_{\odot}$ at a distance of the LMC (taken
to be 55 kpc), for a lens at a distance of 10 kpc, and for a relative
lens-source velocity at the lens of 250 km sec$^{-1}$, $T$ is on the order
of 4 hours.  Note that this corresponds to the distance in time between the
inflection points in Figure 1 and it does not depend on the mass of the
lens. To detect inflection points one must then sample the light curve on a
time-scale significantly shorter than $T$, which is significantly less than
the daily rate of most of the currently running MACHO with the exception of
the EROS CCD program (Aubourg et al. 1993), which has a sampling rate of 22
minutes. The crossing time is also significantly greater than the typical
exposure time for all the MACHO searches.

\bigskip

\cl{\bf 3. Discussion and Summary }

Currently only EROS's CCD program (Aubourg et al. 1993) has the time
resolution necessary to see a situation where a MACHO crosses in front of a
stellar disk. We consider it somewhat unlikely that the present search
techniques of the MACHO collaboration or OGLE will encounter such a
situation. This is because the time between repeated images of the same
stellar field are long compared to the duration of events that stars with
masses less than $10^{-3}$ $\Mo$ are likely to create. Also, the MACHO
collaboration and OGLE do not have the time resolution needed to accurately
discern inflection points near the peaks of light curves. However, if their
search techniques are augmented with a search on shorter time intervals,
say repeating some observations on the time scales of hours instead of
days, finite source size effects may become important (Gould 1992, Aubourg
et al. 1993, Witt and Mao 1994). For expected values of $V$, one would
expect this to encompass the mass scale between about $10^{-9}$ and
$10^{-3}$ $\Mo$.

A lens search of source stars in M31 (Crotts 1992) would be sensitive to
smaller mass lenses and hence finite source sizes might be more prominent
and equally valuable in deconvolving lens and source parameters.

Nemiroff (1991) showed that with continuous monitoring, the smaller the
mass of the lens that dominates the Galactic halo, the more frequently lens
events would occur.  This is expected primarily because there are more
lenses needed at lower masses to populate the Galactic halo. These events
would be expected to have increasingly shorter duration, since smaller mass
lenses have smaller Einstein rings but equal spatial velocities, and so
would cross their Einstein rings in a shorter time.  The duration of the
crossing time of the lens in front of the stellar disk, $T$, however, would
be independent of the mass of the lens, since the radius of the source and
the relative velocity of the lens are both independent of this mass.

Modulation effects on microlensing light curves caused by rotation of the
earth would have an effect only if the effective transverse lens velocity
is not large compared to the earth's spin speed, which is about 0.5 km
sec$^{-1}$ at the surface.  Modulation effects by the earths orbital motion
(Gould 1992) would only significantly distort a part of a light curve which
takes on the order of a months, which is significantly longer than current
estimates for $T$.  Rapid binary motion of the lens or source might
significantly change the light curve (Nemiroff 1991, Griest 1991, Gould
1992).

Photometric errors would necessarily create ambiguity in the time of the
inflection points on a microlensing light curve.  This ambiguity would
propagate into $T$ and so into the estimated mass of the lens. Even in the
case of significant errors, however, it might be quite clear that a disk
crossing event is being detected, because the center of the light curve
could be significantly different than expected with a point source (Witt
and Mao 1994).

We describe the advantages of probing the finite source size regime here
because we feel that it is likely such capability is achievable in the next
few years, and it is desirable to call attention to this phenomenon as a
worthy goal.  If, for example, a possible indication of the onset of a
lensing event was taken as a ``trigger," more frequent follow-up
measurements might be made capable of better exploring the center regions
of light curves, where finite source size effects could be dominant.

It is assumed that the source appears circular on the sky and that it has
uniform surface brightness across its face.  These do not appear
unreasonable assumptions for a normal star, especially in the light of the
inaccuracy of the other quantities known. It is possible that bumps in the
light curves between clearly delineated inflection points could give
evidence for star-spots or other non-uniformities in the source star's
appearance  .

In sum, if high magnification lens events could be measured to high
time-precision by MACHO-type search programs, one would expect to measure
lens events where the lens passes directly across the background star's
face.  Such an event would create discernable inflection points in the
lensing light curve which would, given an independent estimate of the
radius of the star, yield an independent estimate of the velocity of the
lens. This lens velocity is additional information that is useful in
determining a more accurate mass of the microlens.

We thank Bohdan Paczynski and Shude Mao for helpful discussions, and Jerry
Bonnell and an anonymous referee for comments and for carefully reading the
manuscript. WADTW acknowledges the support of a Zaccheus Daniel Fellowship.
This work was supported by a grant from NASA.

\bigskip

\cl {\bf REFERENCES }
\parindent=0pt
\baselineskip=14pt
\parskip=6pt

Alcock, C. A. et al. 1993, Nature, 365, 621

Aubourg, E. et al. 1993, Nature, 365, 623

Crotts, A. P. S. 1992, ApJ, 399, L43

Gould, A. 1992, ApJ, 392, 442

Griest, K. 1991, ApJ, 366, 412

Liebes, S. Jr. 1964, Phys. Rev., 133, B835.

Nemiroff, R. J. 1987, Ph D. Thesis, University of Pennsylvania

Nemiroff, R. J. 1991, A\&A, 247, 73

Paczynski, B. 1986, ApJ, 304, 1

Refsdal, S. 1964, MNRAS, 128, 295.

\hangindent=20pt
Schneider, P., Ehlers, J., and Falco, E. E. 1992, Gravitational Lenses,
(Springer-Verlag: Berlin), pp 313-315

Schneider, P. and Wagoner, R. V. 1987, ApJ, 314, 154

Witt J. and Mao, S. 1994, ApJ, submitted.

\hangindent=20pt
Udalski, A., Szymanski, M., Kaluzny, J., Kubiak, M., Krzeminski, W., Mateo,
M., Preston, G. W., and Paczynski, B. 1993, Acta Astron, 43, 289

\vfill\eject
\baselineskip=20pt

\cl{\bf Figure Caption }

\noindent
{\bf Figure 1}: The magnification of a uniform circular disk by a point
lens.  Two independent lens crossings of the disk are shown, each marked by
a long arrow. Position of each lens on the $x$ axis is listed in units of
$E$, the angular Einstein ring size of the lens, with the zero point
defined as the center of the source.  Magnification of the source by each
lens is listed in magnitudes as $\Delta M$.  Note that the times each lens
crosses the disk boundaries are closely matched by inflection points on the
light curve.

\vfill\eject
\end